# Fibonacci Numbers and the Golden Ratio in Biology, Physics, Astrophysics, Chemistry and Technology: A Non-Exhaustive Review

Vladimir Pletser

Technology and Engineering Center for Space Utilization, Chinese Academy of Sciences, Beijing, China; Vladimir.Pletser@csu.ac.cn

**Abstract**

Fibonacci numbers and the golden ratio can be found in nearly all domains of Science, appearing when self-organization processes are at play and/or expressing minimum energy configurations. Several non-exhaustive examples are given in biology (natural and artificial phyllotaxis, genetic code and DNA), physics (hydrogen bonds, chaos, superconductivity), astrophysics (pulsating stars, black holes), chemistry (quasicrystals, protein AB models), and technology (tribology, resistors, quantum computing, quantum phase transitions, photonics).

**1. Introduction**

Fibonacci numbers $F_i = F_{i-1} + F_{i-2}$ with $F_1 = 1$ and $F_2 = 1$ and the golden ratio $\varphi = \lim_{i \to \infty} \frac{F_i}{F_{i-1}} = \frac{1+\sqrt{5}}{2} \approx 1.618\ldots$ can be found in nearly all domains of Science appearing when self-organization processes are at play and/or expressing minimum energy configurations. Several, but by far non-exhaustive, examples are given in biology, physics, astrophysics, chemistry and technology.

## 2. Biology

2.1 <u>Natural and artificial phyllotaxis</u>

Several authors (see e.g. Onderdonk, 1970; Mitchison, 1977, and references therein) described the phyllotaxis arrangement of leaves on plant stems in such a way that the overall vertical configuration of leaves is optimized to receive rain water, sunshine and air, according to a golden angle function of Fibonacci numbers. However, this view is not uniformly accepted (see Coxeter, 1961).

N. Rivier and his collaborators (2016) modelized natural phyllotaxis by the tiling by Voronoi cells of spiral lattices formed by points placed regularly on a generative spiral. Locally, neighboring cells are organized as three whorls or parastichies, labelled with successive Fibonacci numbers. The structure is encoded as the sequence of the shapes (number of sides) of the successive Voronoi cells on the generative spiral.

Fibonacci spiral patterns were produced artificially (Li Chaorong et al., 2005, 2007) by manipulating the stress on inorganic microstructures made of a silver core and a silicon dioxide shell. It was found that an elastically mismatched bi-layer structure may cause stress patterns that give rise to Fibonacci spirals. Results suggest that plant patterns might be modelled by mutually repulsive entities for both spherical and conical surfaces. It is conjectured that Fibonacci spirals are the least energy configuration on conical supports.

Levitov (1991) gave a phyllotaxis model in which Fibonacci sequences and the golden ratio appear in the pattern of spines on a cactus and can be replicated for cylindrically constrained, repulsive objects. A "magnetic cactus" has been developed (Nisoli et al., 2009) with 50 outward-pointing magnets acting as spines, verifying Levitov's model with magnets self-organizing into configurations involving Fibonacci sequences and the golden ratio. Similar dynamics could occur in crystallized ion beams, in which the system self-organizes into concentric cylindrical shells. Dynamical phyllotaxis could also occur in other physical systems such as Wigner crystals in curved nanostructures and trapped ions in cylindrical potentials

2.2 Genetic code and DNA

The Fibonacci sequence, and its "quantum" extension, can be found in genetic codes, including amino acids and codons (Négadi, 2015).

Deoxyribonucleic acid (DNA) in biological systems replicates with the aid of proteins. However, Kim et al. (2015) have designed a controllable self-replicating system that does not require proteins. The self-assembly process into rings continues through two different replication pathways: one grows exponentially, the other grows according to Fibonacci's sequence.

**3. Physics**

3.1 Hydrogen bonds

The golden ratio appears in atomic physics, more specifically in the Bohr radius and bond-valence parameters of hydrogen bonds of certain borates (Yu et al., 2006).

3.2 Chaos

In non-linear chaotic dynamics, the Fibonacci sequence appears in the Feigenbaum scaling of the period-doubling cascade to chaos (Linage et al., 2006), applicable in many dynamical systems (turbulent flows, chemical oscillators, cell biology).

3.3 Superconductivity

Otto (2016) obtained results from an empirical relation between the critical temperature of optimum doped superconductors and the mean cationic charge, indicating the fractal character of high critical temperature superconductivity, for which the width of superconducting domains is governed by Fibonacci numbers.

**4. Astrophysics**

4.1 Pulsating stars

Lindner and his collaborators (2015) have shown, using light curves of the Kepler space telescope, how the brightness of some stars pulsates at primary and secondary frequencies whose ratios are near the golden ratio. A nonlinear dynamical system driven by an irrational ratio of frequencies generically exhibits a strange but nonchaotic attractor.

4.2 Black holes

Black holes warp space in their vicinity so much that in classical General Relativity, nothing, not even light, can escape. However, when quantum effects are included, black holes can lose energy via a process known as Hawking radiation. Black holes are characterized by their mass and their angular momentum. Spinning Kerr black holes can exist in two states: one in which they heat up when losing energy (negative specific heat), and one in which they cool down (positive specific heat). They can also transition from one state into the other. Davies (1977, 1978, 1989) considered the change in internal energy per change in temperature while the ratio of angular momentum $J$ to mass $M$ is held constant and he obtained that the ratio $J^2/M^4$ (in units where the speed of light and gravitational constant are unity) is the inverse of the golden ratio (see also Livio, 2012; Baez, 2013). Other relations between black hole physics and the golden ratio can be found in (Cruz et al., 2917; Nieto, 2011).

**5. Chemistry**

5.1 Quasicrystals

Discovered by Shechtman in 1984 (that earned him the 2011 Chemistry Nobel Prize), quasiperiodic crystals are ordered but not periodic structures (Shechtman et al., 1984; Cahn et al., 1986a, 1986b; Shechtman, 1988). Quasicrystals represent minimum energy structures in shape of polyhedra. When bombarded with gamma rays, these polyhedra produce a diffraction pattern involving the golden ratio. Spectrum displays two quadrupole doublet peaks (Swartzendruber et al, 1985). When fitting least-squares regression for these curves, standard deviations are minimum when the intensity ratio is close to the golden ratio, suggesting that spacings between atoms in the quasicrystalline lattice structure is governed by the golden ratio.

Topological properties of Fibonacci quasicrystals are further investigated in (Levy et al., 2016). Wavefunctions and spectrum for quasiperiodic Fibonacci chains are obtained theoretically in (Macé et al., 2016). Properties of Fibonacci lattices with arbitrary spacings are investigated in (Lo Gullo et al., 2016). One-dimensional quasicrystals are self-similar with the self-similarity factor being the golden ratio (Grushina et al., 2008).

Quasi-periodic mesoscopic Fibonacci rings are used to investigate firstly, the quantum mechanical phenomenon of persistent circular charge current, crucial in understanding quantum coherence, (Patra and Maiti, 2016a) and secondly, spin Berry phase associated to the motion of electrons in presence of a spin-orbit field, responsible for the generation of spin current (Patra and Maiti, 2016b).

5.2 <u>Protein AB models</u>

Lee and collaborators (2008) investigated a global structural optimization for off-lattice protein AB models, consisting of hydrophobic and hydrophilic monomers in Fibonacci sequences, in two and three dimensions by conformational space annealing. Results obtained by introducing a shift operator in the internal coordinate system showed improved results on AB models with new low energy conformations. Zhang and Ma (2009) applied a sampling method for computing a partition function based on random walks on thermodynamic variables, to the off-lattice protein AB models, in which cases many low energy states were found in different models. These off-lattice protein AB models use a Fibonacci series sequence protein with chain lengths of 13, 21, 34, 55, and 89. Kolossváry and Bowers (2012) revisited the Fibonacci sequences and find that much lower energy folds exist than previously reported. They found the protein sequence that yields the lowest energy fold amongst all sequences for a given chain length and residue mixture. In particular, for protein models with a binary sequence, the sequence-optimized folds form more compact cores than the lowest energy folds.

**6. Technology**

6.1 <u>Tribology</u>

A study of frictional effects of periodicity in a crystalline lattice (Jeong Young Park et al., 2005) showed that friction along the surface of a quasicrystal in the direction of a periodic geometric configuration is about eight times greater than in the direction where the geometric configuration is aperiodic. Geometric periodicity was confirmed via rows of atoms that formed a Fibonacci sequence.

6.2 <u>Resistors</u>

Srinivasan (1992) showed that the ratio of the effective resistance of an infinite network of identical resistors to the resistance of a constituent resistor is equal to the golden ratio.

6.3 <u>Quantum Computing</u>

Topological quantum computing uses "braids" of particle tracks. Vaezi and Barkeshli (2014) theoretically showed that anyons tunnelling in a double-layer system can transition to an exotic non-Abelian state that contains Fibonacci anyons. Some phases of two-dimensional electron systems support Fibonacci anyons, which are quasiparticles with the property that a collection of $N$ Fibonacci anyons at fixed positions has a Hilbert space of degenerate states of dimension given by the $N$th Fibonacci number. As a result, quantum information can be stored in a collection of

Fibonacci anyons and braiding supplies a universal gate set for quantum computation (Clarke and Nayak, 2015). Vaezi (2014) obtained more efficient non-Abelian quasi-particles, Fibonacci anyons, with a superconducting vortex, at the interface of a 2/3 fractional quantum Hall-superconductor structure. This system went through a phase transition to a Fibonacci state to build fault-tolerant quantum computers.

The combined effects of extended dimensionality and doping are explored in (Soni et al., 2015) by studying ladders composed of coupled chains of interacting itinerant Fibonacci anyons. These itinerant Fibonacci anyons are further investigated (Ayeni et al., 2016) on a one-dimensional chain where no natural notion of braiding arises and also on a two-leg ladder where anyons hop between sites and possibly braid. Nontrivial braids for three Fibonacci anyons can be used to systematically construct entangling gates for two qubits (Carnahan et al., 2015).

Braiding representations in relation with the Fibonacci model for topological quantum computing based on fusion rules for a Majorana fermion, are discussed in (Kauffman and Lomonaco, 2016)

The topological quantum computation Fibonacci-Levin-Wen code, for which every quantum gate can be approximated by braiding Fibonacci anyons, is explored in (Koenig et al, 2010; Bonesteel and DiVincenzo, 2012; Wosnitzka et al., 2015).

More generally, topological orders are patterns of long-range entanglement in many-body ground states, one of them being a Fibonacci bosonic topological order stacking with a fermionic product state (Tian Lan et al., 2015). Fractional Chern insulators are investigated in (Behrmann et al. 2016), with stability and finite-size properties of generalizations of Fibonacci anyon quantum Hall states. Chern numbers of quasicrystalline structures determined by interferometry for finite-length Fibonacci chains can be observed directly in their diffraction pattern (Dareau et al., 2016).

Fibonacci topological quantum field theories in 5 and 10 layers with Fibonacci anyons model are reported in (Neupert et al., 2016).

6.4 Quantum phase transitions

Quantum phase transitions take place between distinct phases of matter at zero temperature. Near the transition point, exotic quantum symmetries can emerge that govern the excitation spectrum of the system. A symmetry described by the *E8* Lie group with a spectrum of 8 particles was long predicted to appear near the critical point of an Ising chain. This system was realized experimentally (Coldea et al., 2010; Levi, 2010) by tuning the quasi-one-dimensional Ising ferromagnet $CoNb_2O_6$ through its critical point using strong transverse magnetic fields. The spin excitations are observed to change character from pairs of kinks in the ordered phase to spin-flips in the paramagnetic phase. Just below the critical field, the spin dynamics shows a fine structure with two sharp modes at low energies, in a ratio that approaches the golden ratio as predicted for the first two meson particles of the *E8* spectrum.

6.5 Photonics

The Fibonacci sequence has been successfully employed in the development of different photonic devices. The focusing and imaging properties of Fibonacci optical elements, e.g., quasicrystals (Negro et al., 2003; Verbin et al., 2015), gratings (Gao et al., 2011; Verma et al., 2014a, 2014b), lenses (Calatayud et al., 2013; Monsoriu et al., 2013; Ferrando et al., 2014), and zone plates (Dai et al., 2012) are studied in (Jie Ke et al., 2015) along with optical zone plates produced by the generalized Fibonacci sequences and their axial focusing properties. Compared with traditional Fresnel zone plates, the generalized Fibonacci zone plates present two axial foci with equal intensity. Generalized Fibonacci photon sieves are reported in (Junyong Zhang et al., 2015) and Fibonacci aperiodic Bragg reflection waveguides in (Fesenkoa and Tuz, 2015)

**7. Conclusion**

Fibonacci numbers and the golden ratio are found in nearly all domains of Science. Several, examples were presented in biology, physics, astrophysics, chemistry and technology. These examples are far from being exhaustive.

Several authors mention as well that configurations involving Fibonacci numbers and the golden ratio correspond to either self-organization processes or minimum energy configurations.